\def\a{\alpha}\def\b{\beta}\def\c{\chi}\def\e{\epsilon}
\def\f{\phi}
\def\l{\lambda}\def\m{\mu}\def\n{\nu}\def\r{\rho}
\def\y{\eta}\def\x{\xi}\def\z{\zeta}

\def\F{\Phi}
\def\O{\Omega}\def\S{\Sigma}

\def\na{\nabla}
\def\inf{\infty}\def\id{\equiv}\def\mo{{-1}}\def\ha{{1\over 2}}
\def\qu{{1\over 4}}

\def\const{{\rm const}}

\def\gmn{g_{\m\n}}\def\mn{{\mu\nu}}  
\def\ds{ds^2=}

\def\af{asymptotically flat }
\def\fe{field equations }\def\bh{black hole }
\def\coo{coordinates }                                   
\def\bhs{black holes }

\def\ssy{spherically symmetric }

\def\dys{dynamical system }\def\cps{critical points }

\def\RN{Reissner-Nordstr\"om }

\def\section#1{\bigskip\noindent{\bf#1}\smallskip}

\font\small = cmr8

\def\PL#1{Phys.\ Lett.\ {\bf#1}} 
 
\def\PR#1{Phys.\ Rev.\ {\bf#1}}\def\CQG#1{Class.\ Quantum Grav.\ {\bf#1}} 
\def\NP#1{Nucl.\ Phys.\ {\bf#1}}

\def\ref#1{\medskip\everypar={\hangindent 2\parindent}#1}
\def\beginref{\begingroup
\bigskip
\centerline{\bf References}
\nobreak\noindent}
\def\endref{\par\endgroup}
\input epsf
\magnification=1200

\def\ef{e^{-2\F}}\def\es{e^{-2q\S/3}}\def\er{e^{2\r}}\def\en{e^{2\n}}
\def\ez{e^{2\z}}\def\ec{e^{2\c}}\def\ey{e^{2\y}}
\def\ess{e^{-2\S}}
\def\tqt{{3+q^2\over3}}\def\ttq{{3\over3+2q^2}}\def\qt{{q^2\over3}}
\def\tdq{3+2q^2}\def\tuq{3+q^2}\def\rp{r_+}\def\rq{r_-}
\def\ttuq{{3\over3+q^2}}\def\qtuq{{q^2\over3+q^2}}\def\qtq{{2q^2\over3+2q^2}}
\def\hyp{hyperboloid }\def\cps{critical points }\def\xinf{$\x\to-\inf$ } 
\def\ab{asymptotic behaviour }
%%%%%%%%%%%%%%%%%%%%%%%%%%%%%
{\nopagenumbers
July 1999\hfill INFNCATH-9908
\vskip120pt
\centerline{\bf Primary scalar hair in dilatonic theories} 
\centerline{\bf with modulus fields}
\vskip40pt
\centerline{\bf S. Mignemi\footnote{$^\dagger$}{\rm  
e-mail: mignemi@cagliari.infn.it}}
\vskip20pt
\centerline{Dipartimento di Matematica, Universit\`a di Cagliari}
\centerline{viale Merello 92, 09123 Cagliari, Italy}
\centerline{and} 
\centerline{INFN, Sezione di Cagliari}
\vskip80pt
\centerline{\bf ABSTRACT}
{
We study the general spherical symmetric solutions of dilaton-modulus gravity
non\-minimally coupled to a Maxwell field, using methods from the theory of
dynamical systems. We show that the solutions can be classified by the mass,
the electric charge, and a third parameter which we argue can be related to
a scalar charge. The global properties of the solutions are discussed.}
\vfil\eject}

\section{Introduction.}
It is known that \bh solutions of charged dilaton-gravity models, as those
arising in effective string theory, present a quite different behaviour from the
\RN solutions of general relativity [1-2]. This is essentially due to the
non-minimal
coupling of the dilaton to the Maxwell field, which spoils the validity of the 
standard no-hair theorems [3] and hence allows for the presence of a non-trivial
dilaton field.
In spite of this, the dilaton charge is not an independent parameter, but is
still a function of the mass
and the electric charge of the \bh and has henceforth sometimes been called
a secondary hair.

In effective four-dimensional string theory, however, further scalar fields are 
present besides the dilaton, as for example the moduli coming from compactification of higher 
dimensions, which are non-minimally coupled to the Maxwell field [4].
The introduction of these fields may change the properties of the black holes.
A simplified model which takes into account one modulus has been studied 
some time ago [5]. It was shown that an exact \ssy\bh solution of the \fe
can be found by
requiring that the dilaton and the modulus are proportional. 
However, this restriction is not necessary, and it would be interesting to
investigate the properties of the most
general \ssy solutions. In general, it is not possible to find these solutions in
analytic form. (The \fe can in fact be cast in the form of a Toda molecule
system of first order differential equations, which is exactly solvable only
in a few special cases). However, the
qualitative behaviour of the solutions and some quantitative results can be
obtained by studying the Toda dynamical system.
In particular, the metric and the scalar fields will necessarily be regular
at all the points of the integral curves except critical points. Consequently,
in order to determine the global properties of the solutions, as the structure of
their horizons and asymptotic regions, it suffices to study their behaviour at
the critical points of the dynamical system. One drawback of this method is
that only the exterior region of the \bh can be studied. The interior may be
however investigated numerically by continuing the solutions beside the horizon.

In this paper we undertake the investigation of the general solutions of the
model introduced in [5] using this approach, and show that in general there
exists a three-parameter family of \af\bh solutions. This result is interesting
because the third parameter can be presumably related to a scalar charge,
giving therefore an example of primary scalar hair.
In addition to these solutions, the model also admits as a limiting case
a two-parameter family of non-\af \bh degenerate
solutions of the kind discussed in [6]. We also discuss the properties of
extremal \bh solutions, which are of great interest in recent developments
of string and membrane theories.

The paper is organized as follows. In section 1 we describe the model and obtain
the dynamical system associated with the field equations. In section 2 we
discuss the exact black hole solutions, obtained for special values of the
parameters. In section 3 we study the dynamical system in its generality,
while in section 4 we discuss the physical properties of its solutions.
\vfil\eject

\section{1. The action and the field equations.}
We study the action [5]:
$$S=\int d^4x\sqrt{-g}\left[R-2(\na\F)^2-{2\over3}(\na\S)^2-
(\ef+\l^2\es)F^2\right]\eqno(1.1)$$
where $\F$ and $\S$ are the 4-dimensional dilaton and modulus respectively,
$F$ is the Maxwell field strength, and $q$ and $\l$ are coupling parameters.
This action has been obtained by dimensional reduction of heterotic string
effective action [7], with the addition of a non-minimal coupling term
for the modulus, arising from integrating out heavy modes [4].

The \fe ensuing from (1.1) are
$$\eqalign{&R_\mn=2\na_\m\F\na_\n\F+{2\over3}\na_\m\S\na_\n\S+2(\ef+\l^2\es)
\left(F_\m^{\ \r}F_{\n\r}-\qu F^2\gmn\right)\cr
&\na^\m[(\ef+\l^2\es)F_\mn]=0\cr
&\na^2\F=-\ha\ef F^2\cr
&\na^2\S=-{q\l^2\over2}\es F^2\cr}\eqno(1.2)$$
A \ssy solution can be found with Maxwell field strength
$$F_{mn}=Q\e_{mn}\qquad m,n=2,3\eqno(1.3a)$$
and a metric of the form [1]
$$\ds\en(-dt^2+e^{4\r}d\x^2)+\er d\O^2\eqno(1.3b)$$
where $\n$, $\r$, $\F$ and $\S$ are functions of the "radial" coordinate $\x$.
Defining a new function $\z=\n+\r$, the \fe (2) take the simpler form
$$\eqalignno{\z''&=\ez&(1.4a)\cr
\F''&=-Q_1^2e^{2\n-2\F}&(1.4b)\cr
\S''&=-qQ_2^2e^{2\n-2q\S/3}&(1.4c)\cr
\n''&=Q_1^2e^{2\n-2\F}+Q_2^2e^{2\n-2q\S/3}&(1.4d)\cr}$$
with $Q_1^2=Q^2$ and $Q_2^2=\l^2Q^2$, subject to the constraint
$$\z'^2-\n'^2-\F'^2-{1\over3}\S'^2+Q_1^2e^{2\n-2\F}+Q_2^2e^{2\n-2q\S/3}-\ez=0
\eqno(1.5)$$

A first integral of eq.(1.4a) is given by
$$\z'^2=\ez+a^2$$
where $a^2$ is an integration constant, which has been chosen to be non-negative
because otherwise one would obtain solutions with no asymptotic region,
which are not of interest to us. For the moment we consider only strictly
positive values of $a$. As we shall see, the limit $a\to0$, corresponds to
extremal solutions.
Integrating again, with a suitable choice of the origin of $\x$, one gets
$$e^\z={2ae^{a\x}\over 1-e^{2a\x}}\eqno(1.6)$$
where $a$ can be chosen to be positive without loss of generality.
Moreover, from the remaining eqs. (1.4), one obtains the relation
$${1\over q}\S''+\F''+\n''=0$$
which can be integrated, to read
$$\S'=-q(\n'+\F'+c)\eqno(1.7)$$
with $c$ an integration constant.
In view of (1.4) and (1.7), defining
$$\c=\n-\F\qquad\y=\n-{q\over3}\S\eqno(1.8)$$
the \fe can be put in the "Toda molecule" form
$$\eqalign{\c''=&2Q_1^2\ec+Q_2^2\ey\cr
\y''=&Q_1^2\ec+\tqt Q_2^2\ey\cr}\eqno(1.9)$$

In terms of $\c$ and $\y$, the derivatives of the fields $\F$, $\S$ and $\n$
are given by
$$\eqalign{\F'=&\ttq\left(\y'-\tqt\c'-\qt c\right)\cr
\S'=&{3q\over3+2q^2}(\c'-2\y'-c)\cr
\n'=&\ttq\left(\y'+\qt\c'-\qt c\right)\cr}\eqno(1.10)$$
and eq.(1.5) can be written
$$a^2-\ttq\left[\tqt\c'^2+2\y'^2-2\y'\c'+\qt c^2\right]+Q_1^2\ec+Q_2^2\ey=0
\eqno(1.11)$$

The equations (1.9) with the constraint (1.11) can be solved exactly in a few
special cases, which are reported in the following section.

In the general case, they can be recast in the form of a 3-dimensional system of
first-order differential equations. If we define the variables
$$X=\c',\qquad Y=\y',\qquad Z=|Q_2|e^\y$$
then the constraint (1.10) can be considered as a definition of $\ec$.
Eliminating the term $\ec$ from eqs. (1.9), one obtains the system:
$$\eqalign{X'=&Z^2+2P(X,Y,Z)\cr Y'=&\tqt Z^2+P(X,Y,Z)\cr Z'=&YZ\cr}
\eqno(1.12)$$
where
$$P(X,Y,Z)=Q_2^2\ec={1\over\tdq}\left[(\tuq)X^2+6Y^2-6XY-3B\right]-Z^2
\eqno(1.13)$$
with $B={\tdq\over3}a^2-\qt c^2$.

\section{2. Exact solutions.}
{\noindent\it A. The $Q_2=0$ case.}
\smallskip
This limit case corresponds to minimal coupling of $\S$, i.e. $\l\to 0$.
By the no-hair theorem,
the regular solutions should have constant $\S$, as we shall verify.
When $Q_2=0$, the equations (1.9) take the form
$$\c''=2Q_1^2\ec,\qquad\qquad\y''=Q_1^2\ec\eqno(2.1)$$
The first equation can be integrated to give
$$\c'^2=2Q_1^2\ec+b^2\eqno(2.2)$$
with $b$ an integration constant. Moreover, comparing the two equations (2.1),
$$\y'=\ha(\c'-k)\eqno(2.3)$$
with $k$ an arbitrary constant. The constraint equation (1.11) becomes then
$$a^2-{b^2\over2}-\ttq\left[\ha k^2+\qt c^2\right]=0\eqno(2.4)$$
Integrating again (2.2), one gets
$$Q_1e^\c={\sqrt2bAe^{b\x}\over 1-A^2e^{2b\x}}\eqno(2.5)$$
with $A$ an integration constant.

From these results and the relations (1.10), one can now write down the 
general solution in
terms of the physical fields. Rather than giving all the explicit expressions,
let us first consider the "radial" metric function $e^\r=e^{\z-\n}$.
As $\x\to 0$, $e^\r\to\inf$, and hence one can identify this limit with 
spatial infinity.
As $\x\to-\inf$, instead, one has from (1.6), (1.10) and (2.5),
$$e^\r\sim\const\times\exp\left[\left(a-{b\over2}+\ttq\left(\ha k+\qt c\right)
\right)\x\right]\eqno(2.6)$$
which implies that for $\x\to-\inf$, $e^\r\to 0$, giving rise to a singularity,
except in the special case when the constant factor in the exponential vanishes,
in which case $e^\r\to\const$ as $\x\to-\inf$.
In conjunction with (2.4), this request singles out a unique real solution for
the parameters, given by $a=b=-c=-k$.

In order to analyze the metric, it is useful to write it in a 
Schwarzschild-like form, by introducing a new radial coordinate $r$, 
such that $dr=\ez d\x$. In the new \coo,
$$\ds-\en dt^2+e^{-2\n}dr^2+\er d\O^2\eqno(2.7)$$
where the metric functions are now viewed as functions of $r$.
With a suitable choice of the origin of $r$, one has then
$$e^{2a\x}={r-\rp\over r-\rq},\qquad\ez=(r-\rp)(r-\rq),\qquad
1-A^2e^{2a\x}=(1-A^2){r\over r-\rq}\eqno(2.8)$$
with $\rp=2a/(1-A^2)$, $\rq=2aA^2/(1-A^2)$.
Moreover, if one chooses $A$ such that $Q_1=2aA/(1-A^2)$, the physical fields
read, in terms of the new radial coordinate,
$$\eqalign{\en&=1-{\rp\over r}\qquad\qquad\er=r^2\left(1-{\rq\over r}\right)\cr
\ef&=1-{\rq\over r}\qquad\qquad\ess=\const\cr}\eqno(2.9)$$
This is nothing but the well-known GHS solution [1-2]. It describes \af \bhs
with mass $\rp/2$ and charge $Q_1^2=\rp\rq/2$. The surface $r=\rp$ is a
horizon while the point $r=\rq$ is a singularity.

Qualitatively different solutions arise in the special case $A=1$. In this case,
$\ez\sim\ec$, and choosing the origin of $r$ such that $\rp=2a$, one gets
$$\eqalign{\en&=r-\rp\qquad\qquad\er=r\cr
\ef&=r\qquad\qquad\ess=\const\cr}\eqno(2.10)$$
This solution is not \af, but still possesses a horizon at $\rp$ and is 
singular at the origin. It has been investigated in detail in ref. [6].

Another important limit is reached when $a=0$ and corresponds to extremal black
holes with $\rq=\rp$. In fact, in that case $\ez=\x^{-2}$, and proceeding as
before one can show that the existence of a regular horizon implies that also
the parameters $b$, $c$ and $k$ must vanish.
Hence, one has $\ec=(\x+A)^{-2}$, with $A$ an integration constant.
Defining a new coordinate $r=A(1-A\x^\mo)$, 
the metric functions can be finally cast in the form (2.9), with $\rp=\rq=A$.
\bigskip

{\noindent\it B. The $Q_1=0$ case.}
\smallskip
This case corresponds to minimal coupling of $\F$ and can be considered the
limit of (1.1) for $\l\to\inf$. By the no-hair theorem,
the regular solutions must have constant $\F$.
The \fe are now
$$\c''=Q_2^2\ey,\qquad\qquad\y''=\tqt Q_2^2\ey\eqno(2.11)$$
Proceeding as before, one gets
$$\eqalign{Q_2e^\y&=\sqrt\ttuq{2bAe^{b\x}\over 1-A^2e^{2b\x}}\cr
\c'&=\ttuq(\y'-k)\cr}\eqno(2.12)$$
where $b$, $A$, $k$ are integration constants, together with the constraint 
$$a^2-\ttuq b^2-\ttq\left[\ttuq k^2+\qt c^2\right]=0\eqno(2.13)$$

We look again for regular \bh solutions. For this purpose, we consider the asymptotic
behaviour of $e^\r$ as $\x\to-\inf$, which is now
$$e^\r\sim\const\times\exp\left[\left(a-\ttuq b+\ttq\left(\qtuq k+\qt c
\right)\right)\x\right]\eqno(2.14)$$
A horizon can only occur when the coefficient of $\x$ in the exponential 
vanishes, in which case $e^\r\to\const$ as $\x\to-\inf$.
This condition, together with (2.13) implies that $a=b=-c=-{3k/q^2}$.

Defining a new coordinate $r$ as before, for $A$ such that $Q_2=\sqrt\ttuq
{2aA\over1-A^2}$, one gets finally
$$\eqalign{\en&=\left(1-{\rp\over r}\right)\left(1-{\rq\over r}\right)
^{3-q^2\over\tuq}\qquad\qquad
\er=r^2\left(1-{\rq\over r}\right)^{2q^2\over\tuq}\cr
\ef&=\const\qquad\qquad\qquad\qquad
\ess=\left(1-{\rq\over r}\right)^{6q\over\tuq}\cr}\eqno(2.15)$$
These solutions have not been considered previously, but essentially coincide
with the generalized GHS solutions[1-2], where now $\S$ plays the role of the
dilaton. They describe \af \bhs
with mass $\ha\left(\rp+{3-q^2\over3+q^2}\rq\right)$ and charge $Q_2^2=\ttuq
\rp\rq$. A horizon occurs at $r=\rp$ and a singularity at $r=\rq$. The extremal
limit $\rp=\rq$ is achieved when $a=b=c=k=0$.

Also the limit $A=1$ is special and describes non-\af
\bhs. For $A=1$ one has, in fact,
$$\eqalign{\en&=(r-\rp)r^{3-q^2\over\tuq}
\qquad\qquad\er=r^{2q^2\over\tuq}\cr
\ef&=\const\qquad\qquad\ess=r^{6q\over\tuq}\cr}\eqno(2.16)$$
Metrics of this form have been investigated in [6].

\bigskip
{\noindent\it C. The case $\c'=\y'$.}
\smallskip
The last case in which exact solutions can be obtained is given by the 
condition $\y=\c+\const$, which corresponds to the solutions found in [5]. 
Setting $\ey=K^2\ec$, the \fe become
$$\c''=(2Q_1^2+K^2Q_2^2)\ec=\left(Q_1^2+\tqt K^2Q_2^2\right)\ec\eqno(2.17)$$
Hence, $K^2={3\over q^2}{Q_1^2\over Q_2^2}={3\over\l^2q^2}$, and
$$Q_1e^\c=\sqrt{q^2\over\tdq}{2bAe^{b\x}\over 1-A^2e^{2b\x}}\eqno(2.18)$$
where $b$ and $A$ are integration constants. The constraint (1.5) reduces to
$$a^2-{\tuq\over\tdq}b^2-{q^2\over\tdq}c^2=0\eqno(2.19)$$
The solution possesses a horizon if
$$a-{\tuq\over\tdq}b+{q^2\over\tdq}c=0\eqno(2.20)$$
From (2.19) and (2.20), one obtains $a=b=-c$.

In terms of the coordinate $r$ defined above, choosing $A$ such that $Q_1=
\sqrt{q^2\over\tdq}{2aA\over1-A^2}$, the metric functions read
$$\eqalign
{\en&=\left(1-{\rp\over r}\right)\left(1-{\rq\over r}\right)^\ttq\qquad\qquad
\er=r^2\left(1-{\rq\over r}\right)^\qtq\cr
\ef&=\left(1-{\rq\over r}\right)^\qtq\qquad\qquad
\ess=(3/q^2\l^2)^{3\over q}\left(1-{\rq\over r}\right)^{6q\over\tdq}\cr}\eqno(2.21)$$
(Notice that we have exchanged the definition of $\rp$ and $\rq$).
These solutions describe \af\bhs of mass $\ha\left(\rp+\ttq\rq\right)$ 
and charge $Q_1^2={q^2\over\tdq}\rp\rq$ [5].
Also in this case the extremal black holes are obtained for vanishing $a$, $b$
and $c$.

In the special case $A=1$, the solutions reduce to
$$\eqalign{\en&=(r-\rp)r^\ttq
\qquad\qquad\er=r^\qtq\cr
\ef&=r^\qtq\qquad\qquad\ess=(3/q^2\l^2)^{3\over q}
r^{6q\over\tuq}\cr}\eqno(2.22)$$
and describe non-\af \bhs.

\section{3. The dynamical system.}
The dynamical system (1.12) is similar to analogous systems studied in
several contexts [8],
but differs from these because the \cps at finite distance lie
on a compact curve.
It is easy to see, in fact, that all the \cps at finite distance are placed
at the intersection between the plane $Z=0$ and the hyperboloid $P=0$, with
$P$ defined in (1.13), which (except in some degenerate cases) is an ellipse.

In particular, the plane $Z=0$ corresponds to the limit $Q_1=0$. The system
is invariant under $Z\to-Z$, but the $Z<0$ half-space is simply a copy of the
positive $Z$ half-space and has no physical significance. Hence, we shall not
consider it in the following.

The hyperboloid $P=0$ contains the trajectories corresponding to the limit
$Q_2=0$. If $B>0$ it is one-sheeted, while it is two-sheeted if $B<0$.
We shall
consider only the former case. It is easy to see, in fact, that when $B<0$,
the hyperboloid does not intersect the plane $Z=0$ and therefore
there are no critical points at finite distance. It follows that the solutions 
of the \dys
are of oscillatory type, and do not lead to reasonable \bh geometries.
Moreover, the physically relevant solutions are those in the exterior of the
\hyp, which corresponds to $|Q_2|e^\c>0$, i.e. to the external region of the
black hole. Finally, we notice that
in the limit $B=0$, the hyperboloid reduces to a cone and the only critical
point at finite distance
is the origin of the coordinates. This limit corresponds to extremal
black hole solutions.

As noted above, when $B>0$, the intersection of the \hyp $P=0$ with 
the plane $Z=0$ is given by an ellipse. More precisely, for every
$$|X_0|\le\sqrt{9B\over(\tdq)(\tuq)}$$
there is a critical point at $X=X_0$, $Y=Y_0$, $Z=0$, where $Y_0$ is given
in terms of $X_0$ by the solution of the quadratic equation
$$(\tuq)X_0^2+6Y_0^2-6X_0Y_0-3B=0\eqno(3.1)$$

The characteristic equation for small perturbations,
$$\eqalign{X=X_0+x,\qquad&|x|\ll1\cr Y=Y_0+y,\qquad&|y|\ll1\cr
Z=Z_0+z,\qquad&|z|\ll1\cr}$$
has eigenvalues $0$, $2X_0$ and $Y_0$.
Hence, each point in the $Z=0$ plane satisfying (3.1) with $X_0>0$, $Y_0>0$,
repels a 2-dimensional bunch of solutions in the full 3-dimensional phase space,
while solutions of (3.1) with $X_0<0$, $Y_0<0$ are attractors.
The points with $X_0>0$, $Y_0<0$ or $X_0<0$, $Y_0>0$ act as saddle points.
The presence of a vanishing eigenvalue is due of course to the fact that
there is a continuous set of critical points lying on a curve.
The critical points correspond to \xinf for trajectories starting from the
ellipse, and to the limit $\x\to\inf$ for trajectories ending at the ellipse.

The trajectories in the $Z=0$ plane, which correspond to the exact
solutions discussed in sec. 2.A,  are given by
lines of equation $Y=\ha(X-k)$, with $k$ a constant. The lines which do not
intersect the ellipse of critical points, i.e. those with $|k|>\sqrt B$,
correspond to
oscillatory behaviour of $X$ and $Y$ and are not of interest to us. Notice
that the extremal trajectories, for which $|k|=\sqrt B$ are tangent to the
ellipse at $X=0$.

The projection of the trajectories corresponding to the exact solutions of
section 2.B from the \hyp to the $Z=0$ plane presents a similar phase portrait.
The trajectories satisfy in this case the equation $Y=\tqt(X-k')$
and only trajectories with $|k'|<\sqrt{B/3}$ cross the ellipse. The
projections of the
extremal trajectories corresponding to $|k'|=\sqrt{B/3}$ are now tangent to
the ellipse at $Y=0$.

For completeness, we notice that the solutions of sect. 2.c are given by the
hyperbola of equation
$$(\tuq)(\tdq)Z^2-3(\tuq)X^2=-9B,\eqno(3.2)$$
lying in the plane $X=Y$.

We pass now to consider the behaviour of the metric functions $e^\r$, $e^\n$,
for \xinf. In this limit, $\ec\sim e^{2X_0\x}$, $\ey\sim e^{2Y_0\x}$, and hence,
$$\eqalign{
e^\r&\sim\exp\left[{1\over\tdq}((\tdq)a+q^2c-q^2X_0-3Y_0)\x\right]\cr
\en&\sim\exp\left[{2\over\tdq}(-q^2c+q^2X_0+3Y_0)\x\right]\cr}\eqno(3.3)$$
In general, the radius $e^\r\to 0$ as $\x\to-\inf$, except in the special case
$$(\tdq)a+q^2c-q^2X_0-3Y_0=0\eqno(3.4)$$
This equation, combined with (3.1), gives the only real solution $X_0=Y_0=a=-c$. In this
case $e^\r\to\const$ as \xinf.
When these conditions are not satisfied, 
the metric function $\en$ is singular near the
critical points, giving rise to a singularity as $e^\r\to 0$.

Also when the relation (3.4) is satisfied, the metric function $\en$ behaves
singularly near the critical points, but this can be shown to be simply
a coordinate singularity by computing the curvature invariants, which tend to a
constant value as \xinf when (3.4) holds.
Therefore, all the trajectories starting from the point $X_0=Y_0=\sqrt
{\ttuq B}$ correspond to solutions with regular horizon, provided $c=-a$.

To complete the analysis of the phase space we must also investigate the 
nature of the critical points on the surface at infinity. This can be done 
by defining new \coo $u$, $y$, and $z$ such that infinity corresponds to
$u\to 0$:
$$u={1\over X},\qquad y={Y\over X},\qquad z={Z\over X}$$
Then eqs. (1.12) take the form
$$\eqalign{\dot u&=-(z^2+2p)u\cr\dot y&=-(z^2+2p)y+\tqt z^2+p\cr
\dot z&=(y-z^2-2p)z\cr}\eqno(3.5)$$
where we have defined $p=P/X^2$ and a dot denotes $u\, d/d\x$.
The critical points with $u=0$ can be classified in three categories:

1) Two critical points, which we denote $L_{1,2}$ placed at $y=1/2$, $z=0$,
i.e.
$$X=\pm\inf,\qquad Y={X\over2},\qquad Z=0$$
These are the endpoints of the trajectories lying in the $Z=0$ plane.
The analysis of stability shows that the point with $X>0$ (resp. $X<0$) acts 
as an attractor (resp. repellor) both on the trajectories coming from finite 
distance and on the two-dimensional bunch of trajectories lying on the 
surface at infinity.

2) Two critical points $M_{1,2}$ lie at $y=(\tuq)/3$, $z^2=(\tuq)/3$, i.e.
$$X=\pm\inf,\qquad Y=\tqt X,\qquad Z=\sqrt\tqt X$$
These are the endpoints of the trajectories lying on the hyperboloid $P=0$.
The analysis of stability shows that also in this case
the point with $X>0$ (resp. $X<0$) attracts (resp. repels) both the 
trajectories coming from finite distance and those lying on the 
surface at infinity.

3) Two critical points $N_{1,2}$ lie at $y=1$, $z^2=3/(\tdq)$, i.e.
$$X=\pm\inf,\qquad Y=X,\qquad Z=\sqrt\ttq X$$
These are the endpoints of the hyperbola (3.2) in the $X=Y$ plane. 
The points with $X>0$ (resp. $X<0$) act as attractors (resp. repellors) on 
the trajectories coming from finite distance and as saddle points on the
trajectories at infinity.

\epsfysize=10truecm\epsfbox{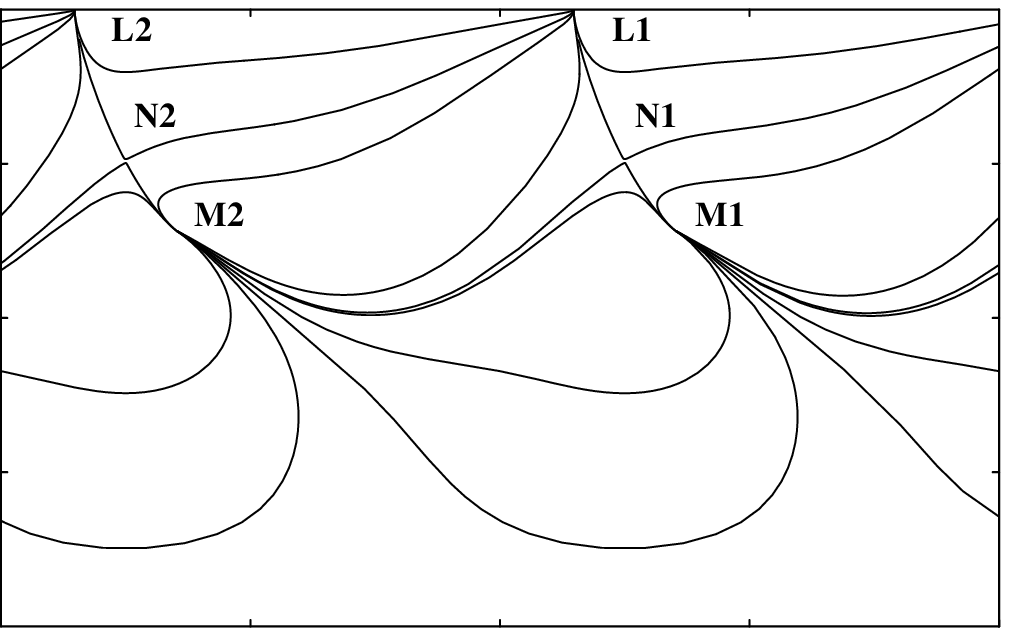}

{\noindent\small Fig. 1: The phase space at infinity.}
\bigskip

In fig. 1 we sketch the pattern of trajectories on the surface at infinity.
The point at infinity is reached for $\x\to\x_0$, where $\x_0$ is a finite
constant. It is easy to see that for $\x\to\x_0$, the functions $\c$ and $\y$ 
behave as
$$e^\c\sim|\x-\x_0|^{1/v_0}\qquad\qquad e^\y\sim|\x-\x_0|^{y_0/v_0}$$
where $v_0\id z_0^2+2p_0$, the subscript $0$ indicating the value taken at the
critical points. Hence, if $\x_0\ne0$, for $\x\to\x_0$, the metric functions
behave as
$$e^\n\sim e^{-\r}\sim|\x-\x_0|^{\ttq(y_0+\qt)v_0^\mo}\eqno(3.6)$$
$$e^\F\sim|\x-\x_0|^{\ttq(y_0-\tqt)v_0^\mo}\qquad
e^\S\sim|\x-\x_0|^{{3q\over\tdq}(1-2y_0)v_0^\mo}$$
The following picture of the phase space emerges:
a family of trajectories start at the ellipse and end at one of 
the \cps $L_1$, $M_1$, $N_1$. Another family of trajectories start
at one of the \cps  $L_2$, $M_2$ or $N_2$ and end at the ellipse.
Moreover, there are trajectories which never intersect the ellipse,
connecting the critical points at $X=-\inf$ to those at $X=+\inf$.
Of all the trajectories, only those starting at points of the ellipse
such that $X_0=Y_0$ can correspond to regular solutions.

For completeness, we observe that most of the trajectories lying in the
interior of the hyperboloid join $M_1$ to $M_2$, but we shall not study
them in detail because they are devoid of physical significance.

\section{4. Discussion.}
We finally discuss the implications of the phase space portrait of the previous
section on the physical properties of the solutions. For this purpose, it is
useful to define a new radial coordinate $r$ such that $dr=\ez d\x$, as in sec. 2.
One has:
$$r={\rp-\rq e^{2a\x}\over 1-e^{2a\x}}\eqno(4.1)$$
where we have defined 
$\rp=2a(1-e^{-2a\x_0})^\mo$, $\rq=2ae^{-2a\x_0}(1-e^{-2a\x_0})^\mo$ .
In this way it is easy to identify the range of variation of $\x$ with the 
corresponding physical regions of the spacetime.

For $\x\to\mp\inf$, $r\to r_\pm$, while for $\x\to0$, $r\to\inf$. Moreover, for 
$\x\to\x_0$, which without loss of generality we shall assume non-negative,
$r\to0$, except when $\x_0$ vanishes.

If $\x_0\ne 0$, we can identify the trajectories starting at the ellipse
and ending at the point $\x_0$ with the exterior region of the
black hole $r>\rp$.
If the condition (3.4) is satisfied, these solutions possess a regular
horizon. Moreover, they are \af, since $\ec$ and $\ey$ tend to a constant 
as $\x\to0$. One can calculate the behaviour of these solutions as $r\to\rp$
From (3.3) and (3.4), one sees that for regular solutions $\en\sim e^{2a\x}$
and hence, $\en\sim(r-\rp)$ for $r\to\rp$. In the same way one can see that
the scalar fields are constant in that limit.
It may be noticed that eq. (3.6) implies that in the unphysical
limit $r\to0$, $\en\sim e^{-2\r}$.

With our conventions, the trajectories starting at $X=-\inf$ and ending at the
ellipse correspond to the unphysical region $0<r<\rq$.
Unfortunately, since $\rq$ is in general a
singularity, one cannot single out the trajectories corresponding to physical
solutions by requiring the regularity of the curvature invariants near that 
point, as for $\rp$.
Moreover, with our methods, we are not able to connect the solutions
in the region $r>\rp$ with those in $r<\rp$ and then to discuss their 
behaviour at $r=\rq$. This may however be achieved by using numerical methods.

The case $\x_0=0$ needs a separate discussion. The solutions are no longer
\af, but their behaviour for $r\to\inf$ can be
obtained from the $\x\to0$ limit, which in our case turns out to be
$$e^\n\sim|\x|^{\ttq(y_0+\qt)v_0^\mo}\qquad
e^\r\sim|\x|^{1-\ttq(y_0+\qt)v_0^\mo}$$
$$e^\F\sim|\x|^{\ttq(y_0-\tqt)v_0^\mo}\qquad
e^\S\sim|\x|^{{3q\over\tdq}(1-2y_0)v_0^\mo}$$
Moreover, since for $\x\to0$, $r\sim|\x|^\mo$, it follows that for $r\to\inf$
the solutions behave in one of the following three ways,
depending on the critical points where they terminate,
\bigskip
\halign{$\quad#\qquad$&$\qquad#$\hfil&$\qquad#$\hfil&$\qquad#$\hfil&$\qquad#$\hfil\cr
&\en&\er&\ef&\ess\cr
L_{1,2}&r&r&r&\const\cr
M_{1,2}&r^{6/(\tuq)}&r^{2q^2/(\tuq)}&\const&r^{6q/(\tuq)}\cr     
N_{1,2}&r^{(6+2q^2)/(\tdq)}&r^{2q^2/(\tdq)}&r^{2q^2/(\tdq)}&r^{6q/(\tdq)}\cr}
\bigskip\noindent
These patterns coincide with those of the exact solutions (2.10), (2.16) or (2.22):
hence all solutions of the
system (1.2) are either \af or possess the same asymptotic behaviour as one of
the exact non-flat solutions. Moreover, from the discussion of sect. 3 of the
phase space at infinity, it
follows that the points $N_{1,2}$ are unstable, so that most trajectories
of this class actually behave like the solutions (2.10) or (2.16) for $r\to\inf$.

As noticed above, the other relevant limit case, $B=0$, corresponds to the
extremal \bh limit. In this case,
the only critical point at finite distance is the origin of coordinates, and
all the eigenvalues of the linearized equations vanish. This degeneration
corresponds to a power-law behaviour of the variables $X$, $Y$ and $Z$
near the critical point: $X\sim -\a\x^\mo$, $y\sim -\b\x^\mo$, $Z\sim \x^{-\b}$,
$\sqrt P\sim \x^{-\a}$. One can easily see from the \fe that the only possible values for
$\a$ and $\b$ are $\a=1, \b=\ha$, $\a=1, \b=1$, and $\a={3\over 3+q^2}, \b=1$,
which coincide with those of the exact extremal solutions of section 2. One can also
check numerically that only the values $\a=1, \b=1$ are stable, so that all the
trajectories, except the exact ones, behave near the critical point at the
origin like the solutions (2.21) (case C). This is interesting, because from
the previous
discussion we know that this limit corresponds to the horizon of the extremal
black hole. Now, it is well known that in the cases A and
C of section 2, the extremal "string" metric $d\hat s^2=e^{2\f}ds^2$ has a "near-horizon"
limit in which the metric function $\er$ becomes constant [9], and hence the
spacetime decouples in the direct product of two 2-dimensional spaces, 
while this is not true for case B. 
But since all solutions except A and B, behave like C near the horizon,
we can conclude that solution B is the only one for which $\er$ is not constant
near the horizon.

Before concluding this section, it is important to remark
that the qualitative properties of the phase space
and hence of the solutions are unaffected by the value of the parameter $q$,
which is therefore essentially irrelevant for our discussion.

\section{5. Conclusions.}
From the previous discussion results that there is a large class of \af regular \bh
solutions of the \fe (1.2). These are characterized by
three parameters: mass, electric charge (or equivalently $\rp$ and $\rq$, or
$a$ and $\x_0$), and a third parameter which classifies the different 
trajectories starting from the critical points $X_0=Y_0=\sqrt{\ttuq B}$,
$Z_0=0$.
We conjecture that the third parameter can be related to (a combination of)
the scalar charges of the dilaton and the modulus.
This conjecture cannot be checked explicitly because only in
a few special cases the solution can be written in an analytic form.

The presence of an independent scalar charge would represent a novelty in the
context of the no-hair results. In fact, in the known cases of dilaton gravity
with non-minimal dilaton-Maxwell coupling, even if the dilaton is non-trivial,
its charge is not an independent parameter, but is related to the mass and 
electric charge of the \bh (secondary hair). In our case of two non-minimally
coupled scalar fields, it seems instead that a new independent charge is needed 
in order to classify the solutions.

Another interesting result is that in the extremal limit all the solutions 
except the unphysical case of a minimally coupled dilaton, have
the same behaviour near the horizon, decoupling into the product of two
2-dimensional spaces. This is interesting since such a behaviour is required
in recent attempts of calculating black hole entropy by counting microstates
of a conformal field theory [10].

Finally, we have clarified the role of non-\af solutions, which were first
discussed in [6] in the case of ordinary dilaton gravity, and shown that in our
model they form a two-parameter family whose \ab can assume only three
possible forms.

\centerline{\bf Acknowledgements}
\smallskip
I wish to thank David Wiltshire for interesting discussions and Massimiliano
Porcu for help with the numerical calculations.
This work was partially supported by a coordinate research project of the
University of Cagliari.

\beginref
\ref[1] G.W. Gibbons, K. Maeda, \NP{B 298}, 741 (1988);
\ref[2] D. Garfinkle, G.T. Horowitz and A. Strominger, \PR{D 43}, 3140 (1991);
\ref[3] J.D. Bekenstein, \PR{D 5}, 1239 (1972); \PR{D 5}, 2403 (1972);
\ref[4] V. Kaplunowsky, \NP{B 307}, 145 (1988);
J. Dixon, V. Kaplunowsky and J. Louis, \NP{B 355}, 649 (1991);
\ref[5] M. Cadoni, S. Mignemi, \PR{D 48}, 5536 (1993);
\ref[6] K.C.K. Chan, J.H. Horne and R. Mann, \NP{B 447}, 441 (1995);
\ref[7] E. Witten, \PL{B 155}, 151 (1985);
\ref[8] D.L. Wiltshire, \PR{D 36}, 1634 (1987); \PR{D 44}, 1100 (1991);
S. Mignemi and D.L. Wiltshire, \CQG{6}, 987 (1989);
\ref[9] S.B. Giddings and A. Strominger, \PR{D 46}, 627 (1992);
M. Cadoni and S. Mignemi, \NP{B 427}, 669 (1994);
\ref[10] See for example J. Maldacena, J. Michelson and A. Strominger,
JHEP {\bf 9902}, 011 (1999) and references therein.

\endref
\end